\documentstyle[sprocl]{article}

\input{epsf}

\def\Journal#1#2#3#4{{#1} {\bf #2}, #3 (#4)}

\def\NPB{{\em Nucl.\ Phys.}\ B}
\def\NPPS{{\em Nucl.\ Phys.\ Proc.\ Suppl.}\ }
\def\PLB{{\em Phys.\ Lett.}\ B}
\def\PL{{\em Phys.\ Lett.}\ }
\def\PRL{{\em Phys.\ Rev.\ Lett.}\ }
\def\PRD{{\em Phys.\ Rev.}\ D}

\newcommand{\md}{\mbox{d}}
\newcommand{\beq}{\begin{equation}}
\newcommand{\eeq}{\end{equation}}
\newcommand{\bea}{\begin{eqnarray}}
\newcommand{\eea}{\end{eqnarray}}

\def\simgt{\rlap{\lower 3.5 pt \hbox{$\mathchar \sim$}} 
                                \raise 1pt \hbox {$>$}}
\def\simlt{\rlap{\lower 3.5 pt \hbox{$\mathchar \sim$}} 
                                \raise 1pt \hbox {$<$}}
\def\sz{\scriptsize}

\def\be{\begin{equation}}
\def\ee{\end{equation}}
\def\bea{\begin{eqnarray}}
\def\eea{\end{eqnarray}}

\begin{document}

\begin{flushright}
 CERN-TH/98-406\\
 December 1998
\end{flushright}

\vspace*{1cm}

\title{HIGHER-ORDER CORRECTIONS TO RADIATIVE\\ $\Upsilon$ DECAYS\footnote{
Talk presented at the IVth International Symposium on Radiative
Corrections (RADCOR 98), Barcelona, September 8-12, 1998, to appear in
the proceedings.}}

\author{MICHAEL KR\"AMER}

\address{Theoretical Physics Division, CERN,\\ 
1211 Geneva 23, Switzerland\\E-mail: Michael.Kraemer@cern.ch} 

\maketitle
\abstracts{Recent advances in the theoretical description of radiative 
$\Upsilon$ decays are reviewed, including the calculation of
next-to-leading order QCD corrections to the photon spectrum.}

\section{Introduction}
The calculations of heavy quarkonium decay rates are among the
earliest applications of perturbative QCD~\cite{AP-75} and have been
used to extract information on the QCD coupling at scales of the order
of the heavy-quark mass. A consistent and rigorous framework for
treating inclusive quarkonium decays has recently been developed,
superseding the earlier non-relativistic potential model.  The
factorization approach is based on the use of non-relativistic QCD
(NRQCD) to separate the short-distance physics of heavy-quark
annihilation from the long-distance physics of bound-state
dynamics.\cite{BBL-95} The annihilation decay rate can be expressed as
a sum of terms, each of which factors into a short-distance
coefficient and a long-distance matrix element:
\begin{equation}\label{eq:fac}
\Gamma(H\to X) = \sum_n C(Q\overline{Q}[n]\to X) 
                 \langle H | {\cal O}_n | H \rangle .
\end{equation}
The sum extends over all possible local 4-fermion operators ${\cal
O}_n$ in the effective Lagrangian of NRQCD that annihilate and create
a heavy-quark--antiquark pair. The short-distance coefficients $C$ are
proportional to the rates for an on-shell $Q\overline{Q}$ pair in a
colour, spin and angular-momentum configuration $n$ to annihilate into
a given final state $X$, and can be calculated perturbatively in the
strong coupling $\alpha_s(m_Q)$. The long-distance factors on the
other hand are well-defined matrix elements that can be evaluated
numerically using lattice simulations of NRQCD.\cite{BSK-96} For small
average velocities $v$ of the heavy quark in the quarkonium rest
frame, each of the NRQCD matrix elements scales with a definite power
of $v$ and the general expression of Eq.~(\ref{eq:fac}) can be
organized into an expansion in powers of the heavy-quark velocity.

At leading order in the velocity expansion, the NRQCD description of
$S$-wave quarkonium decays is equivalent to the non-relativistic
potential model, where the non-perturbative dynamics is described by a
single long-distance factor related to the bound state's wave function
at the origin.\footnote{In the case of $P$-wave quarkonia, the
potential model calculations at ${\cal O}(\alpha_s^3)$ encounter
infrared divergences, which can not be factored into a single
non-perturbative quantity. Within the NRQCD approach, this problem
finds its natural solution since the infrared singularities are
factored into a long-distance matrix element related to higher
Fock-state components of the quarkonium wave function.} The NRQCD
approach improves upon the potential model calculations for $S$-wave
quarkonia by providing a rigorous non-perturbative definition of the
long-distance factor, so this can be calculated using lattice
simulations, and by allowing the systematic inclusion of relativistic
corrections due to the motion of the heavy quarks in the bound state.

Using the NRQCD factorization approach it is, in principle, possible
to calculate the annihilation decay rates of heavy quarkonium from
first principles, the only inputs being the heavy-quark mass $m_Q$ and
the QCD coupling $\alpha_s(m_Q)$. Fixing the heavy-quark mass from,
for example, sum rule calculations for quarkonia, the analysis of
quarkonium decay rates can provide useful information on the QCD
coupling at the heavy-quark mass scale.

For a reliable extraction of $\alpha_s(m_Q)$, however, the theoretical
uncertainties in the calculation of quarkonium decay rates have to be
analysed carefully, and the effects of higher-order corrections in
both the perturbative expansion and the velocity expansion have to be
included. Radiative $\Upsilon$ decays are particularly well suited to
study the impact of higher-order corrections, since not only the total
decay rate but also the photon spectrum can be measured and compared
with theory.

\section{The leading-order decay rate $\Upsilon \to \gamma + X$}
\begin{figure}[htb]
   \vspace*{-3cm}
   \hspace*{0cm}
   \epsfysize=20cm
   \epsfxsize=16cm
   \centerline{\epsffile{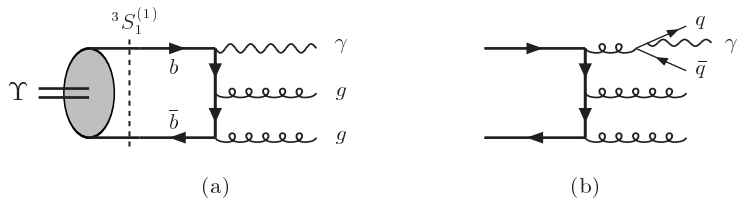}}
   \vspace*{-14.5cm}
\caption[dummy]{\small \label{fig:feyn-lo} Generic leading order Feynman 
diagrams for radiative $\Upsilon$ decay: (a) direct contribution, (b)
fragmentation contribution.}
\end{figure}

At leading order in the velocity expansion, radiative $\Upsilon$
decays proceed through the annihilation of a colour-singlet
$n={}^{3}S_1^{(1)}$ $b\bar{b}$ pair. (We use spectroscopic notation
with the superscript in brackets denoting the colour state.) The direct
contribution, where the photon is radiated off a heavy quark, is shown
in Fig.~\ref{fig:feyn-lo}(a).
The corresponding leading-order decay rate is given by
\begin{equation}\label{eq:rate-lo}
\Gamma_{\mbox{\sz dir}}^{\mbox{\sz LO}}(\Upsilon\to \gamma X) = 
\frac{16}{27}\,\alpha\,\alpha_s^2\, e_b^2
\,(\pi^2-9)\,\frac{\langle \Upsilon | {\cal O}_1({}^3S_1) | \Upsilon 
\rangle}{m_b^2} ,
\end{equation}
where $e_b$ is the charge of the $b$ quark.\cite{CHAN-75} Up to
corrections of ${\cal O}(v^4)$, the non-perturbative NRQCD matrix
element is related to the $\Upsilon$ wave function at the origin
through $\langle \Upsilon | {\cal O}_1({}^3S_1) | \Upsilon \rangle
\approx 2 N_C \,|\, \varphi(0)\,|^2$ with $N_C=3$ the number of colours. 
More information about the decay dynamics is provided by the shape of
the photon energy spectrum, which, at leading order in the velocity
expansion and for the direct term, can be predicted within
perturbation theory
\begin{eqnarray}\label{eq:spec-lo}
\frac{1}{\Gamma_{\mbox{\sz dir}}^{\mbox{\sz LO}}}
\frac{\md\Gamma_{\mbox{\sz dir}}^{\mbox{\sz LO}}(\Upsilon\to 
\gamma X)}{\md x_\gamma} &=& 
\frac{2}{(\pi^2-9)}\left(
2\,\frac{1-x_\gamma}{x_\gamma^2}\,\ln(1-x_\gamma)\right.\nonumber\\
&&\hspace*{-1cm}\left. -2\,\frac{(1-x_\gamma)^2}{(2-x_\gamma)^3}\,
\ln(1-x_\gamma) +\frac{2-x_\gamma}{x_\gamma}
+x_\gamma\,\frac{1-x_\gamma}{(2-x_\gamma)^2}\right) ,
\end{eqnarray}
where $0\leq x_\gamma = E_\gamma/m_b \leq 1$.\cite{BGHC-78} To very
good accuracy, Eq.~(\ref{eq:spec-lo}) can be approximated by a linear
spectrum: $1/\Gamma_{\mbox{\sz dir}}^{\mbox{\sz LO}} \, \md
\Gamma_{\mbox{\sz dir}}^{\mbox{\sz LO}}/\md x_\gamma \approx 2 x_\gamma$.

\begin{figure}[htb]
   \vspace*{-5mm}
   \epsfysize=6cm
   \epsfxsize=8cm
   \centerline{\epsffile{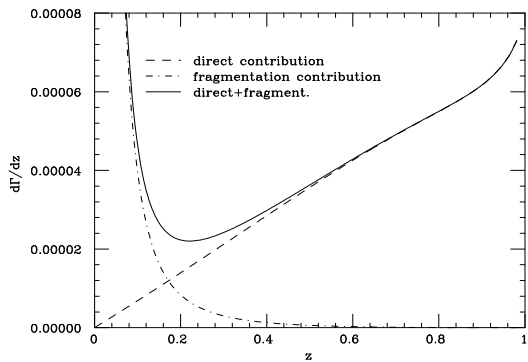}}
   \vspace*{-1cm}
\caption[dummy]{\small \label{fig:frag} The leading-order photon spectrum 
$\Upsilon\to\gamma X$.\cite{CH-95} Shown are the direct component
(dashed curve), the fragmentation contribution (dot-dashed curve) and
the sum (solid curve). Units $\langle \Upsilon | {\cal O}_1({}^3S_1) |
\Upsilon \rangle / m_b^2 = 6/\pi$ have been used, and $\alpha_s = 
0.2$.}
\end{figure}

It has been pointed out recently that there is an additional
leading-order contribution to radiative $\Upsilon$ decays, where the
photon is produced by fragmentation, i.e.\ by collinear emission from
light quarks, see Fig.~\ref{fig:feyn-lo}(b).\cite{CH-95} The
fragmentation contribution, although of ${\cal O}(\alpha\alpha_s^4)$
in the coupling constant, is enhanced by a double logarithmic
singularity $\sim \ln^2(m_b^2/\Lambda^2)\sim 1/\alpha_s^2$ arising
from the phase-space region where both the light-quark--antiquark
splitting as well as the photon emission become collinear. The
complete LO photon spectrum is thus given by
\beq 
 \frac{\md\Gamma^{\mbox{\sz LO}}(\Upsilon\to \gamma X)}{\md x_\gamma} 
= \frac{\md\Gamma_{\mbox{\sz dir}}^{\mbox{\sz LO}}}{\md x_\gamma} 
+ \int_{x_\gamma}^1\,\frac{\md{z}}{z}\,C_{g}(z)\, D_g^{\gamma}
\left(\frac{x_\gamma}{z}\right) ,
\eeq
where the first term on the right-hand side denotes the direct cross
section, Eq.~(\ref{eq:spec-lo}), while the second term represents the
contribution from gluon fragmentation into a photon.\cite{CH-95} The
fragmentation function $D_g^{\gamma}$ is sensitive to non-perturbative
effects and has to be extracted from experiment. The relative size of
the direct component and the fragmentation contribution to the photon
spectrum has been studied at leading order and is shown in
Fig.~\ref{fig:frag}.
The LO fragmentation contribution is important in the low-$x_\gamma$
region, but suppressed with respect to the direct cross section for
$x_\gamma\;\simgt\;0.3$.\footnote{At next-to-leading order, the
fragmentation contribution involves also quark fragmentation, which
is much harder than gluon fragmentation and may be important at larger
values of $x_\gamma$.}

It is important to point out that the total decay width, including
fragmentation processes, $\Gamma(\Upsilon\to\gamma X) = \int_0^1 \,
\md x_\gamma \, \md \Gamma / \md x_\gamma $, is not an infrared-safe
observable. For $x_\gamma\to 0$, the fragmentation contribution rises
like $1/x_\gamma$, characteristic of the soft bremsstrahlung spectrum,
and cannot be integrated down to $x_\gamma =0$. Therefore, in
perturbation theory, only the photon energy spectrum can be calculated
for $x_\gamma \neq 0$, after collinear singularities have been
factorized into the fragmentation functions.

\begin{figure}[htb]
   \vspace{0cm}
   \epsfysize=6cm
   \epsfxsize=7cm
   \centerline{\epsffile{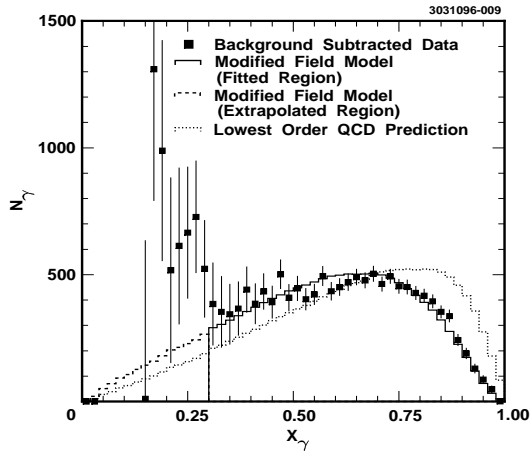}}
   \vspace*{-2mm}
\caption[dummy]{\small \label{fig:cleo} Experimental data for the photon 
spectrum compared to the efficiency-attenuated and energy-smeared
prediction for the direct process at leading order.\cite{CLEO-97}}
   \vspace*{-5mm}
\end{figure}

At present, only the region $x_\gamma \;\simgt\;0.3$ is accessible to
accurate measurement, because of the strong background from
$\pi\to\gamma\gamma$ decays at low $x_\gamma$. A comparison between
the theoretical prediction for the shape of the photon energy spectrum
and the available experimental data can thus be restricted to the
direct term.  Such a comparison is shown in
Fig.~\ref{fig:cleo}.\cite{CLEO-97}
The theoretical curve has been modified to account for the
experimental efficiency and energy resolution and has been normalized
to the data.  The steep rise towards large $x_\gamma$ predicted by LO
perturbative QCD is not supported by the experimental data. In
particular for $x_\gamma\;\simgt\; 0.7$ the discrepancy is
significant. Near the phase-space boundary $x_\gamma \to 1$,
hadronization effects may become important, but they are not expected
to modify the energy distribution significantly in the intermediate
region $x_\gamma\sim 0.7$. To accommodate the data, different models
have been proposed, including a parton shower Monte Carlo approach,
also shown in Fig.~\ref{fig:cleo}, or parametrizations of
non-perturbative effects in terms of an effective gluon
mass.\cite{FIELD-83,CF-94} A proper theoretical understanding of the
photon spectrum, however, requires a consistent inclusion of
higher-order terms in the perturbative expansion as well as a careful
examination of the contributions from relativistic corrections.

Given the above-mentioned theoretical problems, the present
extractions of $\alpha_s(m_b)$ from the total radiative decay rate
have to be considered uncertain. Even neglecting fragmentation
contributions at low $x_\gamma$, model approaches have to be used to
extrapolate the experimental data into the lower energy region to
obtain the total decay rate, introducing uncontrolled theoretical
uncertainties.  Analysing the effect of higher-order corrections to
the photon energy spectrum is thus not only an interesting problem of
perturbative QCD per se, but will also lead to a more reliable
determination of the strong coupling constant from radiative
$\Upsilon$ decays.

\section{Higher-order corrections to $\Upsilon\to\gamma+X$}
The general factorization formula for the $\Upsilon$ decay rate of
Eq.~(\ref{eq:fac}) is a double power-series expansion in the strong
coupling $\alpha_s$ and the heavy-quark velocity $v$. Higher-order
contributions to the LO estimate include both ${\cal O}(\alpha_s)$
corrections to the short-distance cross section and 
relativistic corrections ${\cal O}(v^2)$ due to the motion of the $b$
quarks in the $\Upsilon$ bound state.  Subsequently, we will discuss
the different sources of higher-order corrections and examine their
impact on the photon energy spectrum.

\subsection{Soft-gluon resummation}
It has been argued that potentially large logarithms $\ln(1-x_\gamma)$
associated with the imperfect cancellation between real and virtual
emission of soft gluons as $x_\gamma \to 1$ may contribute to all
orders in perturbation theory.\cite{PHOT-85} The resummation of these
soft-gluon effects may then give rise to a Sudakov suppression $\sim
1/\exp(\alpha_s\ln^2(1-x_\gamma))$ near the endpoint of the photon
energy spectrum. However, a recent analysis reveals that the
logarithms of $(1-x_\gamma)$ cancel at each order in the perturbative
expansion, in the non-relativistic limit where $\Upsilon$ decays
proceed via the annihilation of a colour-singlet $b\bar{b}$
pair.\cite{CHM-00} Accordingly, no Sudakov suppression arises at
leading order in the velocity expansion. It has also been shown that
the parton shower Monte Carlo approach,\cite{FIELD-83} used in the
recent determination of $\alpha_s$ from radiative $\Upsilon$ decays,
does not correctly take into account gluon radiation in higher
orders.\cite{CHM-00}

\subsection{Relativistic corrections}
Although the velocity of the $b$ quarks in the $\Upsilon$ bound state
is small, $v^2 \approx 0.1$, relativistic corrections may contribute
significantly in certain regions of phase space. The NRQCD approach
allows a systematic calculation of these corrections, introducing
however additional non-perturbative matrix elements. Thus, in the
absence of lattice results, one has to resort to phenomenological
inputs to estimate the size of the higher-order terms in the velocity
expansion.

The leading relativistic correction to the colour-singlet decay rate
is of ${\cal O}(v^2)$ and can be expressed in terms of the `binding
energy' of the quarkonium $M_\Upsilon - 2 m_b$.\cite{GK-97} Depending
on the value of the $b$-quark pole mass, the ${\cal O}(v^2)$
correction may be as large as 25\%, but it affects the shape of the
photon spectrum only in the region $x_\gamma\;\simgt\;
0.8$.\cite{KM-83} At ${\cal O}(v^4)$ in the velocity expansion,
radiative $\Upsilon$ decays can proceed through the annihilation of
colour-octet $n={}^{1}S_0^{(8)}, {}^{3}P_{0,2}^{(8)}$ $b\bar{b}$ pairs
into $\gamma g$ final states at ${\cal O}(\alpha\alpha_s)$.  The
enhancement of the colour-octet short-distance annihilation cross
section $\sim \pi/\alpha_s$ partly compensates the strong suppression
of the long-distance matrix element $\sim v^4$. Since the LO
short-distance process for colour-octet terms is kinematically
restricted $\sim\delta(1-x_\gamma)$, colour-octet contributions to the
direct component of the decay spectrum are important only in the
endpoint region $x_\gamma\; \simgt\; 1 - v^2
\approx 0.9$, taking into account an effective smearing of the delta
function due to the resummation of higher-order terms in the velocity
expansion.\cite{RW-97} Colour-octet processes also contribute to
radiative $\Upsilon$ decays by fragmentation.\cite{MP-98} Both gluon
and quark fragmentation turn out to be sizeable at low photon energies
$x_\gamma\;\simlt\; 0.3$. A separate measurement of the gluon
fragmentation functions from the lower part of the energy spectrum, as
suggested by the analysis of the colour-singlet decay channel, may
thus not be feasible.

To summarize, relativistic corrections to the photon energy spectrum,
although potentially sizeable, are concentrated in the upper and lower
endpoint region and do not significantly modify the shape of the
spectrum in the intermediate energy range $0.4 \;\simlt \; x_\gamma
\;\simlt\; 0.9$.

\subsection{Next-to-leading order QCD corrections}
For intermediate photon energies, the spectrum is expected to be well
described by the direct colour-singlet channel. The next-to-leading
order corrections to the total decay rate of Eq.~(\ref{eq:rate-lo})
are potentially large, depending on the choice of the renormalization
scale. An earlier calculation yields
\beq\label{eq:rate-nlo}
\Gamma_{\mbox{\sz dir}}^{\mbox{\sz NLO}}(\Upsilon\to \gamma X) = 
\Gamma_{\mbox{\sz dir}}^{\mbox{\sz LO}}\,
\left(1-\frac{\alpha_s(2m_b)}{\pi}\,(1.67\pm 0.36)\right) , \quad\quad\;
\mu = 2m_b , 
\eeq
with a large theoretical uncertainty of $\sim\pm 20\%$ coming from the
numerical evaluation of the loop integrals.\cite{ML-81} 

\begin{figure}[htb]
   \vspace*{-3cm}
   \hspace*{0cm}
   \epsfysize=20cm
   \epsfxsize=16cm
   \centerline{\epsffile{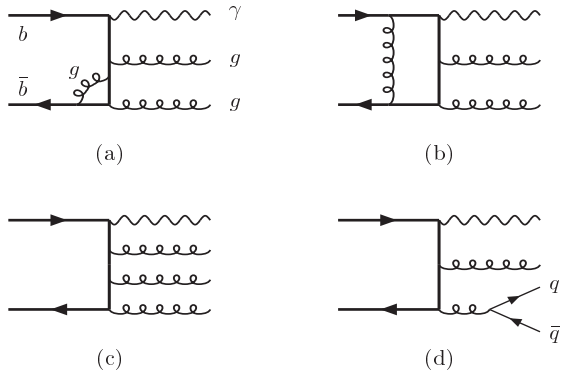}}
   \vspace*{-11.5cm}
\caption[dummy]{\small \label{fig:feyn-nlo} Generic next-to-leading order 
Feynman diagrams for direct radiative $\Upsilon$ decay.}
   \vspace*{-2mm}
\end{figure}

The calculation of the higher-order QCD corrections to the photon
energy spectrum has been completed only recently.\cite{KS-98} Generic
diagrams that build up the decay rate in next-to-leading order are
depicted in Fig.~\ref{fig:feyn-nlo}.
Besides the usual self-energy and vertex corrections for photon and
gluons (a), one encounters a large number of box diagrams (b) as well
as gluon radiation off heavy quarks (c) and gluon splitting into
gluons and light-quark--antiquark pairs (d). The evaluation of these
amplitudes has been performed in the Feynman gauge and dimensional
regularization has been used to calculate the singular parts of the
amplitude. The masses of the $n_{\mbox{\sz lf}} = 4$ light quarks have
been neglected while the mass parameter of the $b$ quark has been
defined on-shell. The exchange of Coulombic gluons in diagram (4b)
leads to a singularity $\sim \pi^2/2\beta_R$, which can be isolated by
introducing a small relative quark velocity $\beta_R$. The
Coulomb-singular part can be associated with the interquark potential
of the bound state and has to be factored into the non-perturbative
NRQCD matrix element. Only the exchange of transverse gluons
contributes to the next-to-leading order expression for the
short-distance annihilation rate. The infrared singularities cancel
when the emission of soft and collinear final-state gluons and light
quarks, described by universal splitting functions~\cite{CS-97}, is
added to the virtual corrections. The analytical result for the matrix
element squared has been implemented in a Monte Carlo integration
program, so that not only the inclusive decay rate but any
distribution can be generated.

For the total width we obtain at next-to-leading order:
\beq\label{eq:rate-nlo2}
\Gamma_{\mbox{\sz dir}}^{\mbox{\sz NLO}}(\Upsilon\to \gamma X) = 
\Gamma_{\mbox{\sz dir}}^{\mbox{\sz LO}}\,
\left[1-\frac{\alpha_s(\mu)}{\pi}\left(1.71
+\beta_0(n_{\mbox{\sz lf}})
\ln\frac{2m_b}{\mu}\right)\right] ,
\eeq
where $\beta_0(n_{\mbox{\sz lf}}) = (11N_C-2n_{\mbox{\sz
lf}})/3$.\cite{KS-98} The theoretical uncertainty due to the numerical
phase-space integration has been reduced to $\simlt\;0.5\%$. The new,
more accurate result for the total width is consistent with the
previous calculation~\cite{ML-81} within the estimated numerical
error.

\begin{figure}[htb]
   \epsfysize=7cm
   \epsfxsize=8cm
   \centerline{\epsffile{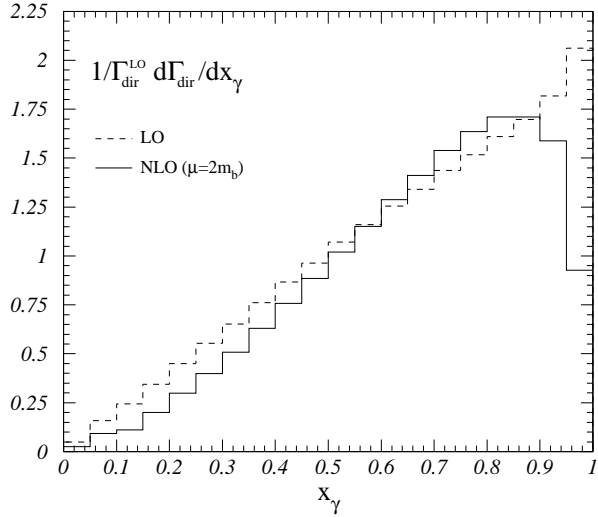}}
   \vspace*{-2.5mm}
\caption[dummy]{\small \label{fig:spec-nlo} Photon energy spectrum 
for direct radiative $\Upsilon$ decays in leading and next-to-leading
order ($\alpha_s = 0.2$).}
\end{figure}

The next-to-leading order corrections significantly flatten and
deplete the photon energy spectrum for $x_\gamma\;\simgt\;0.75$ as
shown in Fig.~\ref{fig:spec-nlo}, indicating that the discrepancy
between theory and experiment at large photon energies can be reduced
by the inclusion of higher-order QCD corrections. A more detailed
discussion of the photon spectrum at NLO and a comparison with
experimental data will be presented in a forthcoming publication.

\section{Conclusions and outlook}
The photon spectrum in radiative $\Upsilon$ decays is a very
interesting laboratory to study perturbative and non-perturbative QCD
effects. In the low-$x_\gamma$ region, fragmentation contributions
dominate, while near the upper endpoint of the spectrum relativistic
corrections and hadronization effects become important. The
next-to-leading order results presented here indicate that the
intermediate region of the energy spectrum can be accounted for by
perturbative QCD, once higher-order corrections are taken into
account. This should allow a more reliable extraction of $\alpha_s$
from radiative $\Upsilon$ decays by restricting the analysis to the
region of the energy spectrum that can be described by NLO
perturbation theory.

\section*{Acknowledgements}
I would like to thank J\"urgen Steegborn for his collaboration during
earlier stages of this work and Johann K\"uhn for his encouragement.
I have benefitted from valuable comments by Stefan Dittmaier and from
discussions with Wim Beenakker, Martin Beneke, Stefano Catani,
Francesco Hautmann, Fabio Maltoni and Michelangelo Mangano. Finally, I
want to thank Joan Sol{\`a} and his colleagues from the Universitat
Aut{\`o}noma de Barcelona for their kind invitation and for organizing
an interesting and enjoyable meeting.  Partial support by DFG under
contract KU 502/7-1 is acknowledged.


\section*{References}
\begin{thebibliography}{999}
\bibitem{AP-75}
 T.~Appelquist and H.D.~Politzer, \Journal{\PRL}{34}{43}{1975}.
\bibitem{BBL-95}
 G.T.~Bodwin, E.~Braaten and G.P.~Lepage, 
 \Journal{\PRD}{51}{1125}{1995}, 
 Erratum, {\em ibid.} \Journal{D}{55}{5853}{1997}.
\bibitem{BSK-96}
 See 
 G.T.~Bodwin, D.K.~Sinclair and S.~Kim,
 \Journal{\PRL}{77}{2376}{1996},
 for first results on NRQCD matrix elements in quenched lattice QCD.
\bibitem{CHAN-75}
 M.S.~Chanowitz, 
 \Journal{\PRD}{12}{918}{1975};   
 T.~Appelquist, A.~De~R\'ujula, H.~Politzer and S.L.~Glashow, 
 \Journal{\PRL}{34}{365}{1975}.
\bibitem{BGHC-78}
 S.J.~Brodsky, T.A.~DeGrand, R.R.~Horgan and D.G.~Coyne, 
 \Journal{\PL}{73B}{203}{1978};   
 K.~Koller and T.~Walsh, 
 \Journal{\NPB}{140}{449}{1978}.   
\bibitem{CH-95}
 S.~Catani and F.~Hautmann, 
 \Journal{\NPPS}{39BC}{359}{1995}.  
\bibitem{CLEO-97}
 CLEO Collaboration, B.~Nemati et al., 
 \Journal{\PRD}{55}{5273}{1997}.
\bibitem{FIELD-83}
 R.D.~Field,
 \Journal{\PL}{133B}{248}{1983}.   
\bibitem{CF-94}
 M.~Consoli and J.H.~Field,
 \Journal{\PRD}{49}{1293}{1994}.
\bibitem{PHOT-85}
 D.~Photiadis,
 \Journal{\PL}{164B}{160}{1985}.   
\bibitem{CHM-00}
 F.~Hautmann,
 to be published in ``Proceedings of the International Conference on the 
 Structure and the Interactions of the Photon'', Egmond aan Zee, 
 Netherlands, 10-15 May 1997 (hep-ph/9708496);
 S.~Catani, F.~Hautmann and M.L.~Mangano, in preparation.
\bibitem{GK-97}
 M.~Gremm and A.~Kapustin, 
 \Journal{\PLB}{407}{323}{1997}.
\bibitem{KM-83}
 W.-Y.~Keung and I.J.~Muzinich,
 \Journal{\PRD}{27}{1518}{1983}.
\bibitem{RW-97}
 I.Z.~Rothstein and M.B.~Wise,
 \Journal{\PLB}{402}{346}{1997}.
\bibitem{MP-98}
 F.~Maltoni and A.~Petrelli, 
 CERN-TH/98-152 (hep-ph/9806455);
 F.~Maltoni, 
 to be published in ``Proceedings of the International Euroconference 
 on Quantum Chromodynamics'', Montpellier, France, 2-8 July 1998 
 (hep-ph/9809260).
\bibitem{ML-81}
 P.B.~Mackenzie and G.P.~Lepage, in ``Proceedings of the Conference on 
 Perturbative QCD'', Tallahassee, USA, 25-28 March, 1981. 
\bibitem{KS-98}
 M.~Kr\"amer and J.~Steegborn, 
 CERN-TH preprint in preparation.
\bibitem{CS-97}
 S.~Catani and M.H.~Seymour, 
 \Journal{\NPB}{485}{291}{1997},
 Erratum, {\em ibid.} \Journal{B}{510}{503}{1997}.
\end{thebibliography}
\end{document}